\documentclass[aip,reprint]{revtex4-2} 
\usepackage{graphicx} 
\usepackage{amsmath} 
\usepackage{hyperref} 
\begin{document}

\title{Supercurrent transmission through Ni/Ru/Ni synthetic antiferromagnets} 
\author{Swapna Sindhu Mishra}
\author{Reza Loloee}
\author{Norman O. Birge}
\email{birge@msu.edu}
\affiliation{Department of Physics and Astronomy, Michigan State University, East Lansing, MI 48824, USA}

\date{\today}

\begin{abstract}
Josephson junctions containing ferromagnetic materials are generating interest for use in superconducting electronics and cryogenic memory. Optimizing the performance of such junctions is an ongoing effort, requiring exploration of a broad range of magnetic systems. Here we study supercurrent transmission through Ni/Ru/Ni synthetic antiferromagnets, with the idea that their magnetic properties may be superior to those of isolated Ni layers. We find that the decay of the supercurrent as a function of Ni thickness is very slow, with a decay length of $7.5\pm0.8$ nm. We also characterize the magnetic properties of the synthetic antiferromagnets as a function of Ni and Ru thicknesses.
\end{abstract}

\maketitle 

Josephson junctions containing ferromagnetic materials are under consideration for applications in cryogenic memory.\cite{Gingrich2016,Dayton2018} The spin-singlet electron pair correlations from the superconducting electrodes undergo rapid phase oscillations and decay in ferromagnetic (F) materials, leading to alternating ``0" or ``$\pi$" junctions depending on the F layer thickness.\cite{Buzdin2005,Ryazanov2001,Kontos2002} In strong F materials the oscillations and decay occur over very short length scales, hence the F layers in S/F/S Josephson junctions (where S=superconductor) must be very thin. Mismatch between the Fermi surfaces of adjacent layers further suppresses the supercurrent. Of the strong F materials studied to date, it has been found that Ni supports the largest supercurrent.\cite{Blum2002,Robinson2006,Baek2018} But thin Ni layers grown on a Nb base electrode behave poorly magnetically at low temperature; in particular the field required to saturate the magnetization grows significantly as the Ni thickness is reduced (see Supplementary Material).

A strategy to improve the magnetic behavior of Ni is to create a Ni/Ru/Ni synthetic antiferromagnet with somewhat different thicknesses of the two Ni layers, $d_{F1} \neq d_{F2}$, called an unbalanced SAF. The idea is that the phase oscillations in the two Ni layers would partially cancel each other, so the SAF would act similarly to a single Ni layer with thickness equal to $d_{F1}-d_{F2}$. The hope is that the thicker Ni layers making up the SAF would have superior magnetic properties than the single thin layer they are replacing. Pursuing this strategy would require extensive device characterization; one doesn't even know if the Ni/Ru interface strongly suppresses supercurrent transmission.\cite{Khasawneh2011}

In this work we take the first step, which is to study supercurrent transmission through balanced Ni SAFs, i.e. with $d_{F1}=d_{F2}$. We first characterize the magnetic properties of the Ni/Ru/Ni SAFs as a function of Ni and Ru thickness, then study supercurrent transmission as a function of Ni thickness. In spite of extensive work in the literature on SAFs based on Co/Cr, Fe/Cr, Co/Ru, NiFe/Cr, NiFe/Ru,  and other systems,\cite{Parkin1990,Donath1991,Belmeguenai2007} we have not found much about the Ni/Ru system. So this work may also be of interest to the magnetism community. 

Thin Ni(2.0)/Ru($d_{Ru}$)/Ni(2.0) films (layer thicknesses in nanometers) were deposited by dc magnetron sputtering on Si substrates in a system with base pressure $4 \times 10^{-6}$ Pa. The Ru thickness, $d_{Ru}$, was varied from 0.6 to 2.7 nm in steps of 0.1 nm near the expected narrow first peak in the coupling strength and 0.2 nm for the wider second peak. A set of Ni($d_{Ni}$)/Ru(0.9)/Ni($d_{Ni}$) films was also sputtered where $d_{Ni}$ was varied from 0.8 to 4 nm in steps of 0.4 nm. All thin film samples have Nb(5)/Cu(2) as the base layer to improve adhesion to the substrate and Cu(2)/Nb(5) as a symmetric capping layer to prevent oxidation. All samples were sputtered at $3.3 \times 10^{-1}$ Pa Ar pressure and at 250 K substrate temperature.

The moment ($M$) vs field ($H$) measurements for both sets of thin films were performed using a SQUID-VSM at 10 K, in fields up to 2 T to ensure that the magnetization was fully saturated.

The Josephson junction fabrication has been discussed in detail elsewhere.\cite{Glick2017} In our junctions we use a [Nb/Al]$_3$/Nb(20) multilayer as our bottom superconducting electrode because it is smoother than pure Nb. The bottom lead shape was patterned using photo-lithography and a lift-off process.  The bottom lead multilayer [Nb(25)/Al(2.4)]$_3$/Nb(20)/Cu(2)/
Ni($d_{Ni}$)/Ru(0.9)/Ni($d_{Ni}$)/Cu(2)/Nb(5)/Au(10) was sputtered where $d_{Ni}$ was varied from 1 to 3 nm. The $\mathrm{1.25 \, \mu m  \times  0.5 \, \mu m}$ elliptical junctions were then patterned by e-beam lithography using a negative ma-N resist and a subtractive process. The surrounding area was ion milled down to the Nb(20) layer and covered with $\mathrm{SiO_x}$ \textit{in-situ} to avoid shorting between the bottom and top superconducting electrodes. After removal of the resist, the top lead shape was patterned using photolithography and another lift-off process. Before deposition of the Nb(150)/Au(10) top superconducting electrode, 5 nm of the Au(10) capping layer was ion milled \textit{in-situ}.

Transport measurements were performed at 4.2 K inside a liquid Helium dewar using a sample probe equipped with a superconducting solenoid. $I$-$V$ curves were measured in fields up to 0.1 T in both directions.

\begin{figure}[!htbp]
\includegraphics[width=\linewidth]{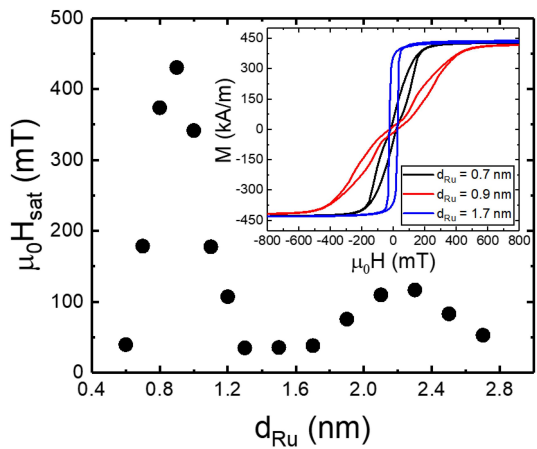}
\caption{Saturation field vs Ru thickness for Ni(2.0)/Ru($d_{Ru}$)/Ni(2.0) synthetic antiferromagnets at $T = 10$ K. Peaks in $H_{sat}$ at $d_{Ru} = 0.9$ and $2.3$ nm indicate maxima in antiferromagnetic coupling. \textbf{Inset:} Magnetization vs field loops of selected samples in the set, also at $T = 10$ K.}
\label{fig:rkky}
\centering
\end{figure}

To characterize the Ru thickness dependence of the magnetic coupling in our Ni/Ru/Ni samples, we measured $M$ vs $H$ curves for a large set of samples with fixed Ni thickness of 2.0 nm and varying Ru thickness. The inset to Fig. \ref{fig:rkky} shows a representative subset that illustrates antiferromagnetic, ferromagnetic and intermediate coupling for $d_{Ru}$ = 0.9, 1.7 and 0.7 nm, respectively. The change in slope at low field in the 0.9 nm data may be an indication of spin-flop behavior occurring in regions of the polycrystalline sample where the magnetocrystalline anisotropy favors alignment with the external field. We ignore that region of the data and estimate the saturation field $H_{sat}$ by extrapolating the approximately linear dependence at higher fields to the saturation value. For samples with the same ferromagnetic layer thickness $d_F$ and saturation moment $M_{sat}$, $H_{sat}$ is a reliable indicator of the coupling strength.

The plot of $H_{sat}$ vs $d_{Ru}$ in Fig. \ref{fig:rkky} shows the first and second peak of the RKKY oscillation to be near Ru thicknesses of 0.9 nm and 2.3 nm respectively. Using the formula for the coupling strength,\cite{Hartmann2013} $J_1 = \mu_0 H_{sat} M_{sat} d_F/2$, we calculate $J_1$ to be 0.18 mJ/m$^2$ (or erg/cm$^2$) and 0.052 mJ/m$^2$ at the first and second peaks, respectively. The coupling strength at the first peak is much weaker than in Co/Ru/Co SAFs but comparable to that in Co/Cu/Co SAFs.\cite{Parkin1991}

\begin{figure}[!htbp]
\includegraphics[width=\linewidth]{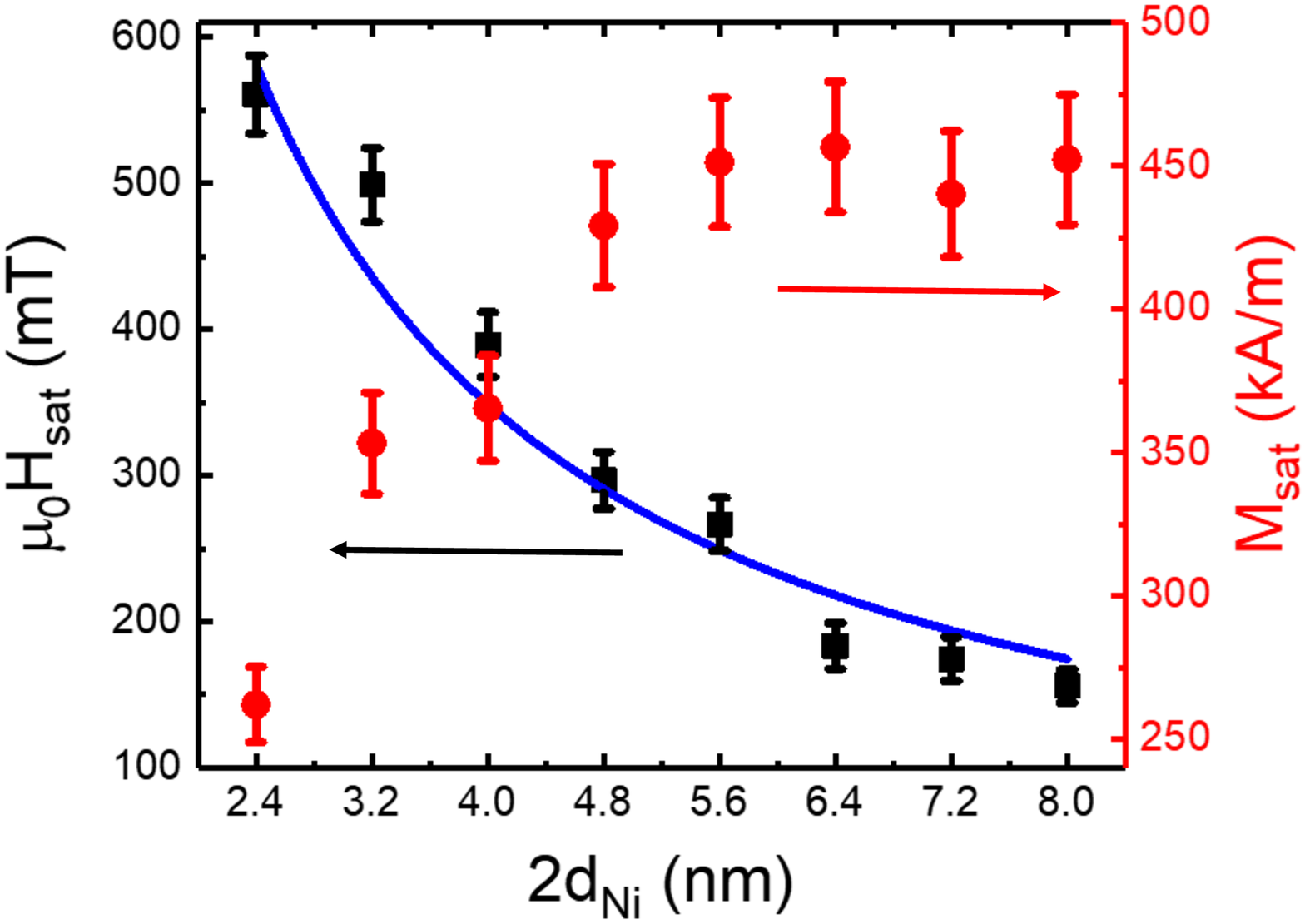}
\caption{Saturation field (black squares, left axis) and saturation magnetization (red circles, right axis) vs total Ni thickness for Ni($d_{Ni}$)/Ru(0.9)/Ni($d_{Ni}$) SAFs measured at 10 K. The blue curve is a fit of $H_{sat}$ to $a/(2d_{Ni})$. Reduction of $M_{sat}$ at small $d_{Ni}$ is an indication of magnetic dead layers.}
\label{fig:saturation}
\centering
\end{figure}

Since we observe the strongest SAF coupling at $d_{Ru} = 0.9$ nm (first peak) in Fig. \ref{fig:rkky}, we use this thickness in all subsequent samples and then varied the Ni thickness from 0.8 to 4.0 nm. All of the samples except the thinnest showed good SAF coupling; the sample with $d_{Ni} = 0.8$ nm showed ferromagnetic coupling. For all of the SAF samples, Fig. \ref{fig:saturation} shows $M_{sat}$ and $H_{sat}$ vs the total Ni thickness, $2d_{Ni}$. The uncertainties $M_{sat}$ are estimated to be 5\% from the area measurement. Uncertainties in $H_{sat}$ were ascribed to the field step size in the measurement.

The blue curve in Fig. \ref{fig:saturation} shows that $H_{sat}$ is inversely proportional to $d_{Ni}$, as expected for a constant interfacial coupling strength. Fig. \ref{fig:saturation} also shows $M_{sat}$, which stabilizes around 450 kA/m for the thick samples -- close to the bulk value of 510 kA/m at low temperature.\cite{Ohandley2000} We attribute the decrease in $M_{sat}$ for the thinner samples to magnetically dead layers at the Ni/Ru and Ni/Cu interfaces. A fit of saturation moment per unit area vs thickness for all the samples indicates that the total magnetic dead layer thickness of the SAFs arising from the two Ni/Cu and two Ni/Ru interfaces is $1.24 \pm 0.07$ nm. (See Supplementary Material.)

\begin{figure}[!htbp]
\includegraphics[width=\linewidth]{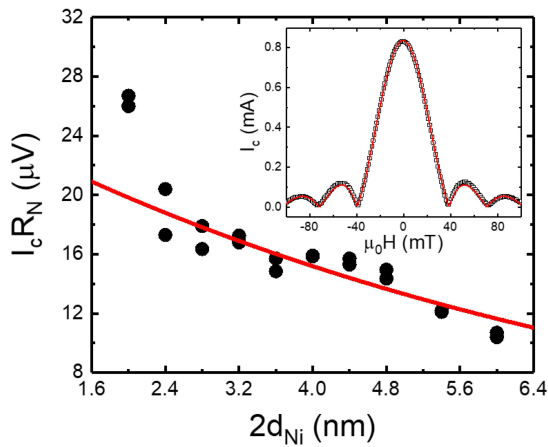}
\caption{Critical current times normal state resistance vs total Ni thickness for Josephson junctions containing Ni/Ru/Ni SAFs. Red curve is a fit to an exponential decay, disregarding the data points at $2d_{Ni} = 2.0$ nm. \textbf{Inset:} Typical critical current vs field data (black squares) fit to the Airy function (red line).}
\label{fig:IcRn}
\centering
\end{figure}

Josephson junctions that contain ferromagnetic materials but do not contain an insulating tunnel barrier generally exhibit overdamped dynamics, with $I-V$ curves that follow the RSJ model:\cite{BaronePaterno1982}
\begin{equation}
    V = \mathrm{sign}(I) R_N Re\left\{\sqrt{I^2-I_c^2}\right\}
\end{equation}
Fits of the model to the $I-V$ data provide estimates of the critical current, $I_c$, and the normal-state resistance, $R_N$.  The inset to Fig. \ref{fig:IcRn} shows the dependence of $I_c$ on magnetic field $H$ applied in the plane of the substrate (perpendicular to the direction of supercurrent flow), which follows an Airy function for elliptical junctions:\cite{BaronePaterno1982}
\begin{equation}\label{eqn:fraunhofer}
    I_c(\Phi) = I_{c,0} \left| \frac{2 J_1 \left( \frac{\pi \Phi}{\Phi_0} \right) }{\frac{\pi \Phi}{\Phi_0}} \right|
\end{equation}
where $I_{c,0}$ is the maximum value of $I_c$, $J_1$ is the Bessel function of first kind, $\Phi_0 = 2 \times 10^{-15}$ Tm$^2$ is the flux quantum, and the total magnetic flux in the system is $\Phi = \mu_0 H (2\lambda_L + d_N + d_F) w$. $\lambda_L$ is the London penetration depth, $d_N$ is the thickness of the normal-metal spacer layers, $d_F$ is the thickness of the ferromagnetic layers, and $w$ is the width of the junctions (measured to be about 550 nm). The Airy pattern is well centered at zero field for both sweep directions (only one of which is shown in the figure), without any visible hysteresis for most of the junctions measured. This implies that the strong antiferromagnetic coupling in the SAF leads to nearly zero remanent magnetization in the junctions.  This does not contradict the hysteresis evident in the magnetization curves shown in the inset to Fig. \ref{fig:rkky}, because those samples were saturated at much higher magnetic fields than the junctions -- see the Supplementary Material. For every junction measured, the maximum value of the Airy function fit, $I_{c,0}$, agrees with the highest measured value of $I_c$ to within 2\%.

We multiply the maximum $I_c$ value by the normal state resistance $R_N$ to obtain $I_c R_N$, which is independent of the junction area. Fig. \ref{fig:IcRn} shows a plot of $I_c R_N$ vs the total Ni thickness, $2d_{Ni}$. We measured two junctions at each thickness to reveal variations in thickness and interface quality during the sputtering process. The junction-to-junction variation in $I_cR_N$ shown in Fig. \ref{fig:IcRn} is typical for Josephson junctions containing thin layers of strong ferromagnetic materials.\cite{Glick2017} The data for the junctions with $d_{Ni}=1.0$ nm are anomalously high compared to the rest of the data. We do not know the reason for that, but it may be related to the ferromagnetic behaviour of the Ni(0.8)/Ru(0.9)/Ni(0.8) film mentioned earlier. We also note that Khasawneh \textit{et al.} observed a similar feature for junctions containing Co/Ru/Co SAFs.\cite{Khasawneh2009}     

The data in Fig. \ref{fig:IcRn} show that $I_cR_N$ decays rather slowly with Ni thickness -- barely a factor of 2 decay as $2d_{Ni}$ ranges from 2.4 nm to 6.0 nm. We have tried fitting the $I_cR_N$ data in Fig. \ref{fig:IcRn} to both an algebraic decay and an exponential decay, but found the former to be unsatisfactory. The exponential fit, shown in the figure, follows the data well if we ignore the first data points at $2d_{Ni} = 2$ nm. From the exponential fit, $I_c R_N = V_0 e^{-2d_{Ni}/\xi_{Ni}}$, we find a value for the decay length to be $\xi_{Ni}=7.5\pm0.8$ nm. If we include the first data points, the decay length decreases to $\xi_{Ni}=5.3\pm0.6$ nm, but the fit does not follow the data as well.

As mentioned earlier, Josephson junctions containing Ni have been studied previously.\cite{Blum2002,Robinson2006,Baek2018} The most comprehensive data were obtained by Baek \textit{et al.}, who studied junctions with Ni thicknesses between 0.9 and 3.9 nm.\cite{Baek2018} Those authors observed minima in $I_cR_N$  corresponding to transitions between the 0-state and the $\pi$-state at Ni thicknesses of about 0.8 and 3.4 nm. $I_cR_N$ reached a maximum value in the $\pi$-state of about 80 $\mu$V or more, for $d_{Ni}\approx 1.8$ nm. The maximum values of $I_cR_N$ in Fig. \ref{fig:IcRn} are several times smaller. That is not surprising: electron transmission through a Ni/Ru/Ni SAF requires good band matching between the majority and minority bands of Ni, since a majority band electron in the first Ni layer becomes a minority band electron in the second layer. In addition, there may be some spin memory loss at the two Ni/Ru interfaces in our SAFs.\cite{Khasawneh2011} Given these two independent effects, it is not possible to determine the effect of either one on our data.

The work presented here is not the first to show supercurrent flow through a SAF. Khasawneh \textit{et al.} \cite{Khasawneh2009} studied Josephson junctions containing Co/Ru/Co SAFs, and also observed an exponential decay of $I_cR_N$ with Co thickness, but with a much shorter decay length, $\xi_{Co}=2.34\pm0.08$ nm. In addition, Robinson \textit{et al.} \cite{Robinson2010} studied Josephson junctions containing Fe/Cr/Fe trilayers, but studied the dependence of $I_cR_N$ only on the dependence of the coupling layer (Cr) thickness, rather than the Fe thickness.

Theoretical calculations for simple S/F/S Josephson junctions (without a SAF) predict that $I_cR_N$ oscillates and decays algebraically with F-layer thickness when the transport is ballistic,\cite{Buzdin1982} or oscillates and decays exponentially when the transport is diffusive.\cite{Buzdin1991} Baek \textit{et al.}\cite{Baek2018} found that their $I_cR_N$ data from Nb/Ni/Nb Josephson junctions was fit well by the ballistic formula for Ni thicknesses ranging from 1 to 4 nm. Substitution of a SAF for the F layer removes the oscillations in $I_c$ due to cancellation of the electron pair phase accumulations in the two F layers of the SAF.\cite{Blanter2004, Vedyayev2005, Crouzy2007, Bakurskiy2015} More remarkably, substitution of a SAF for the F layer in the ballistic limit also removes the decay of $I_c$ with SAF thickness. A Josephson junction containing a perfect SAF with no scattering in the bulk or at the interfaces is predicted to have a value of $I_c$ equal to that of an S/N/S junction of the same thickness.\cite{Blanter2004} That remarkable result is destroyed by the presence of disorder (except for the case of very thin F layers;\cite{Bakurskiy2015}) in the disordered case one expects an exponential decay of $I_c$ with thickness of the SAF. 

The theoretical predictions discussed above are based on a simplified model of a ferromagnet with parabolic bands that are displaced in energy by twice the exchange energy, $E_{ex}$. In addition, it is generally assumed that $E_{ex}$ is much smaller than the Fermi energy, so that differences in Fermi velocities between the majority and minority spin bands can be ignored. Those assumptions are not appropriate for strong ferromagnetic materials such as Ni, Fe, or Co.  Recently, Ness \textit{et al.}\cite{Ness2021} have carried out a calculation of supercurrent through Josephson junctions containing Ni, while incorporating a realistic band structure for the Ni. Those authors used a combination of density functional theory and Bogoliubov-de-Gennes theory to calculate the supercurrent through the Nb/Ni/Nb model junctions. The model junctions contain no disorder; nevertheless, the authors find that the supercurrent oscillates and decays exponentially over the entire thickness range studied, from 0.6 to 12.5 nm. They attribute the exponential decay to two phenomena. At small thicknesses, many of the transverse modes at the Fermi surface occur in only one spin channel, e.g. the majority band, but without a corresponding mode in the minority band. The supercurrent due to such modes decays exponentially over a very short distance. At larger thickesses, supercurrent is carried by modes that propagate over long distances, but which have a sizeable spread in their oscillation periods. Dephasing between such modes leads to an exponential decay of the supercurrent with Ni thickness. Remarkably, Ness \textit{et al.} \cite{Ness2021} find that the thickness dependence of the supercurrent decay is fit well by an oscillation with a single exponential decay constant, equal to $\xi_{Ni}^{calc}=4.1$ nm.

Comparing our own data with the numerical calculation of Ness \textit{et al.} is surprising at first. In our junctions, which certainly contain disorder, we measured a decay length that is \textit{longer} than the value calculated by Ness \textit{et al.} for ideal junctions that contain no disorder. But of course, our junctions contain SAFs. While the most evident consequence of the SAF is the phase cancellation that removes the oscillations in $I_c$, our data suggest a second consequence that is far less obvious -- namely that the decay length of $I_c$ in the SAF is longer than in a single Ni layer. That result is not predicted by the standard theories in the diffusive limit.\cite{Blanter2004,Crouzy2007} Given that our junctions contain many interfaces, we had always assumed that the transport was diffusive rather than ballistic. But the data presented here suggest that that assumption is not completely valid. We know from previous studies that multilayers deposited in our sputtering system, while polycrystalline, exhibit columnar growth, often with epitaxy between the different materials within a single column or grain.\cite{Geng1999} It is plausible that transport through a single column is partly ballistic.  

In conclusion, Ni/Ru/Ni synthetic antiferromagnets exhibit strong antiferromagnetic coupling for a Ru thickness of 0.9 nm. Josephson junctions containing Ni/Ru/Ni SAFs exhibit a very slow decay of the critical current, and no oscillations, as a function of Ni thickness. While the lack of oscillations was expected, the very slow decay suggests that electron pair transport through the SAF is at least partially ballistic. An unbalanced SAF might be used as a replacement for the thin Ni fixed layer currently used in the spin-valve Josephson junctions in Northrop Grumman's cryogenic memory.\cite{Dayton2018}  While the $I_c R_N$ product of such a SAF is likely to be somewhat smaller than that of a single Ni layer,\cite{Baek2018} the prospect of improved magnetic behavior is worth pursuing in the future.

\section*{Supplementary Material}
The Supplementary Material contains data on the hysteresis and saturation magnetization of Ni($d_{Ni}$) films at 10 K, saturation moment per unit area for Ni($d_{Ni}$)/Ru(0.9)/Ni($d_{Ni}$) films at 10 K, and temperature dependence of hysteresis, saturation magnetization and coercivity of Ni(2.0)/Ru(2.3)/Ni(2.0) and Ni(4.0) films. It also discusses why we do not see a field shift in the plots of $I_c R_N$ vs field (the ``Fraunhofer patterns").

\begin{acknowledgments}
The authors thank V. Aguilar, T.F. Ambrose, R. Klaes, M.G. Loving, A.E Madden, D.L. Miller, N.D. Rizzo, and J.C. Willard for helpful discussions. We also thank D. Edmunds for technical assistance and B. Bi for the use of the W. M. Keck Microfabrication Facility. This research was supported by Northrop Grumman Corporation. 
\end{acknowledgments}

\section*{Author Declarations}
\subsection*{Conflict of interest}
The authors have no conflicts to disclose.

\section*{Data Availability}
The data that support the findings of this study are available from the corresponding author upon reasonable request.

\bibliography{mishra_references}

\end{document}


\title{Supplementary Material: Supercurrent transmission through Ni/Ru/Ni synthetic antiferromagnetss} 
\author{Swapna Sindhu Mishra}
\author{Reza Loloee}
\author{Norman O. Birge}
\email[]{birge@msu.edu}
\affiliation{Department of Physics and Astronomy, Michigan State University, East Lansing, MI 48824, USA}

\date{\today}

\maketitle 

\begin{figure}[!htbp]
\includegraphics[width=\linewidth]{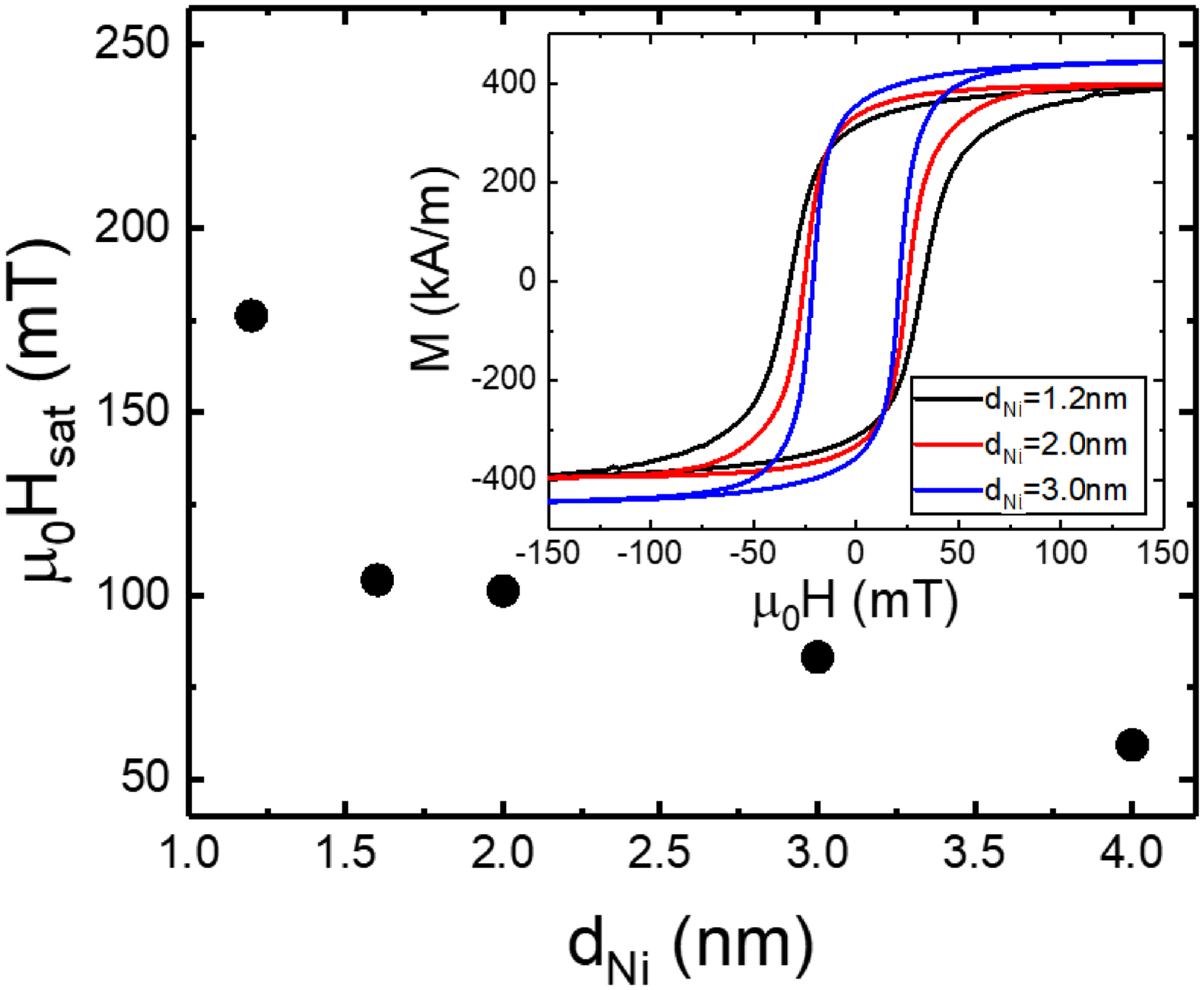}
\caption{Saturation field vs Ni thickness measured at $T=10$ K.}
\label{fig:NiSat}
\centering
\end{figure}

We sputtered Ni($d_{Ni}$) samples with $d_{Ni}$ = 1.2, 1.6, 2.0, 3.0 and 4.0 nm to make various comparisons with the Ni/Ru/Ni SAF samples. As was the case with all of the thin-film samples discussed in the main body, these films were grown on a Nb(5)/Cu(2) base layer, and covered with a Cu(2)/Nb(5) capping layer to prevent oxidation of the Ni. We measured $M$ vs $H$ loops for these samples at 10 K and show a selected set in the inset of Fig. \ref{fig:NiSat}. As evident from the hysteresis loops, the coercivity of Ni grows with decreasing thickness. In addition, the field above which irreversible behavior ceases also grows with decreasing thickness. Since the latter is close to the field where the magnetization saturates, we call it $H_{sat}$, and define it as the field where the magnetization during the upsweep reaches 99\% of the value during the preceding downsweep. Fig. \ref{fig:NiSat} shows the dependence of $H_{sat}$ on the thickness of Ni. At low temperatures, disorder in the form of defects or surface roughness causes $H_{sat}$ to increase rapidly as the film thickness approaches 1 nm, probably due to pinning of domain walls.

\begin{figure}[!htbp]
\includegraphics[width=\linewidth]{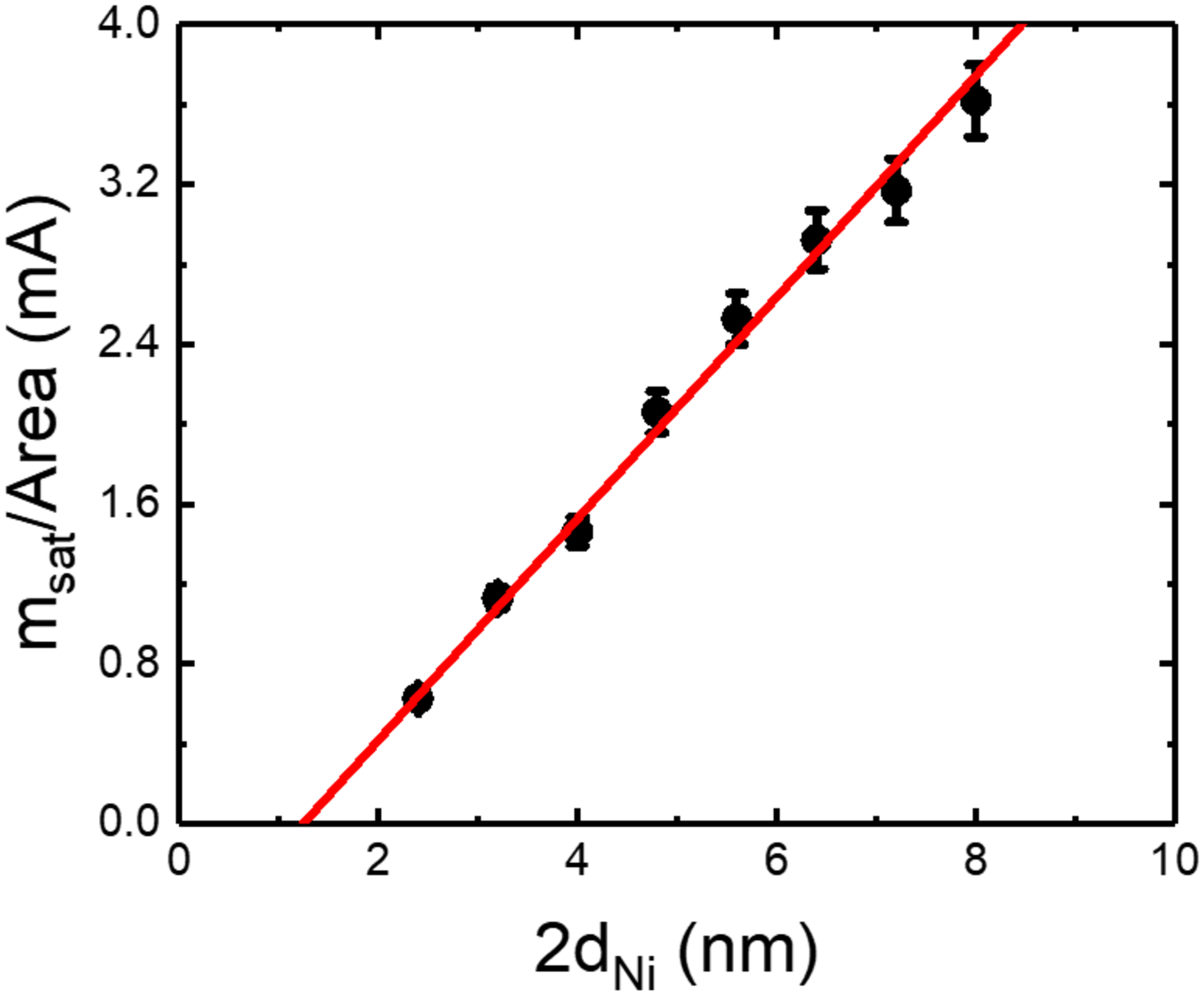}
\caption{Saturation moment per unit area vs total Ni thickness for Ni($d_{Ni}$)/Ru(0.9)/Ni($d_{Ni}$) SAFs measured at 10 K.}
\label{fig:deadlayer}
\centering
\end{figure}

Fig. 2 of the main paper showed a decrease in saturation magnetization of Ni/Ru/Ni SAFs at low Ni thicknesses, which we attributed to magnetic dead layers at the Ni/Cu and Ni/Ru interfaces. Fig. \ref{fig:deadlayer} shows a plot of the saturation moment per unit area vs total Ni thickness for the same Ni($d_{Ni}$)/Ru(0.9)/Ni($d_{Ni}$) SAFs discussed in the main paper. To estimate the total thickness of the dead layers, we fit this plot to a straight line and obtain the x-axis intercept. The total magnetic dead layer thickness is $1.24 \pm 0.07$ nm, which is reasonable given the contributions from two Ni/Cu and two Ni/Ru interfaces. The slope of this fit provides an estimate of the saturation magnetization of Ni, $M_{Ni} = 554 \pm 15$ kA/m. This is a bit higher than bulk, but one should not take this value too seriously because the straight-line fit assumes that the dead layer thickness is constant in all the samples. 

\begin{figure}[!htbp]
\includegraphics[width=\linewidth]{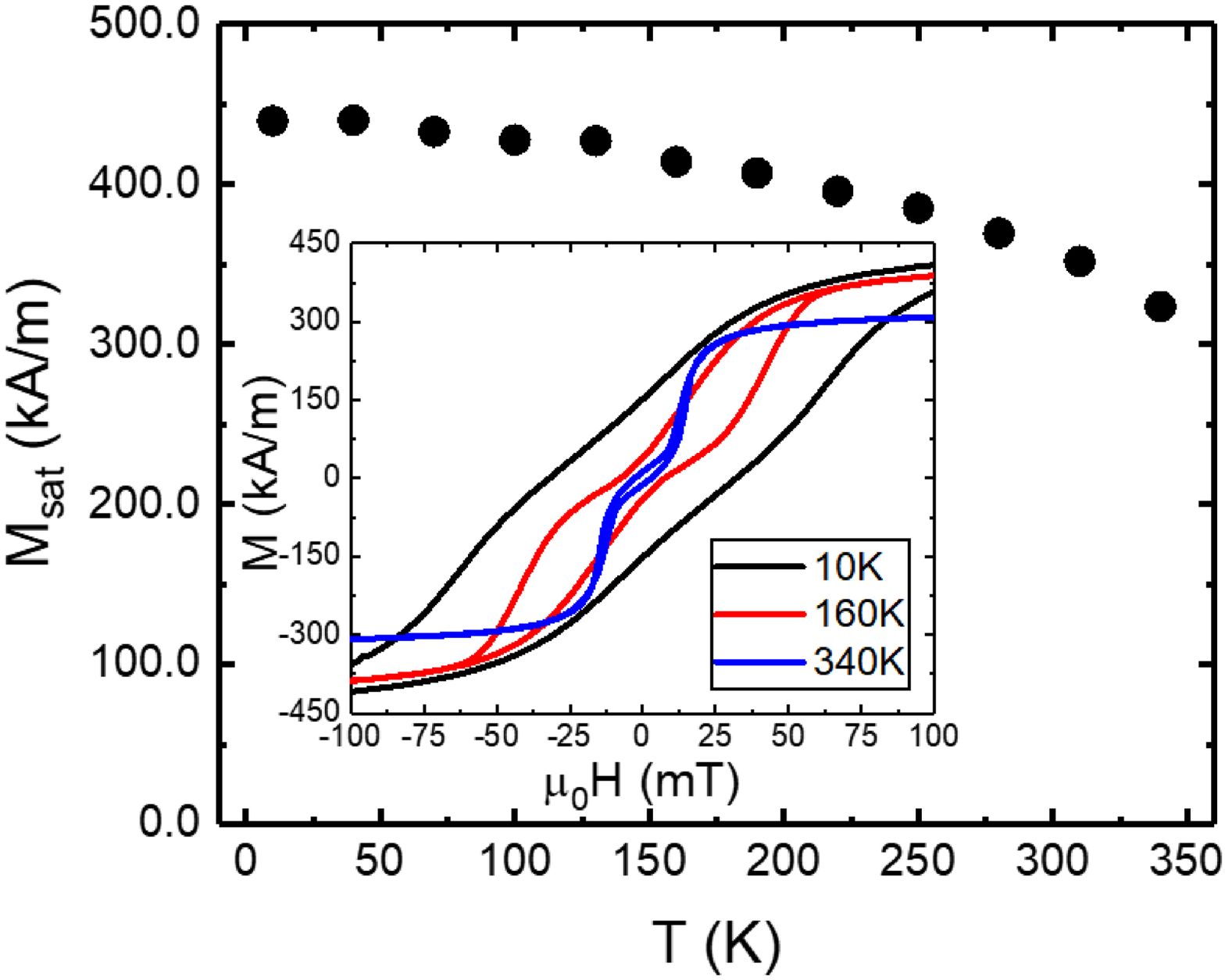}
\caption{Saturation magnetization vs temperature for a Ni(2.0)/Ru(2.3)/Ni(2.0) SAF. \textbf{Inset:} Magnetization vs field loops for selected temperatures. As temperature decreases, the coercivity of the Ni grows, preventing closure of the $M$ vs $H$ loops.}
\label{fig:NiRuTemp}
\centering
\end{figure}

For our Ni(2.0)/Ru($d_{Ru}$)/Ni(2.0) samples we found strong antiferromagnetic coupling near the first peak with $d_{Ru} = 0.9$ nm. The $M$ vs $H$ loops for this Ru thickness show minimal hysteresis. But around the second peak at $d_{Ru} = 2.3$ nm, the loops have wider openings due to the weaker coupling combined with the magnetocrystalline anisotropy and disorder of the thin Ni layers discussed above. We wanted to determine the effect of the low measurement temperature of 10 K on the coupling strength and the shape of the $M$ vs $H$ loops. We measured $M$ vs $H$ loops for a Ni(2.0)/Ru(2.3)/Ni(2.0) SAF from 10 K to 340 K in 30 K steps as shown in Fig. \ref{fig:NiRuTemp}. With increasing temperature, there is a drop in the saturation magnetization value of Ni as expected. However, we also see what appears to be an improvement in the antiferromagnetic coupling between the Ni layers with increasing temperature. As seen in the Fig \ref{fig:NiRuTemp} inset, the hysteretic gap between the loops closes with increasing temperature. We show in the next section that the closing of the hysteretic gap with increasing temperature is actually due to the magnetic behavior of the Ni films, rather than an increase in the antiferromagnetic coupling strength through the Ru.  

\begin{figure}[!htbp]
\includegraphics[width=\linewidth]{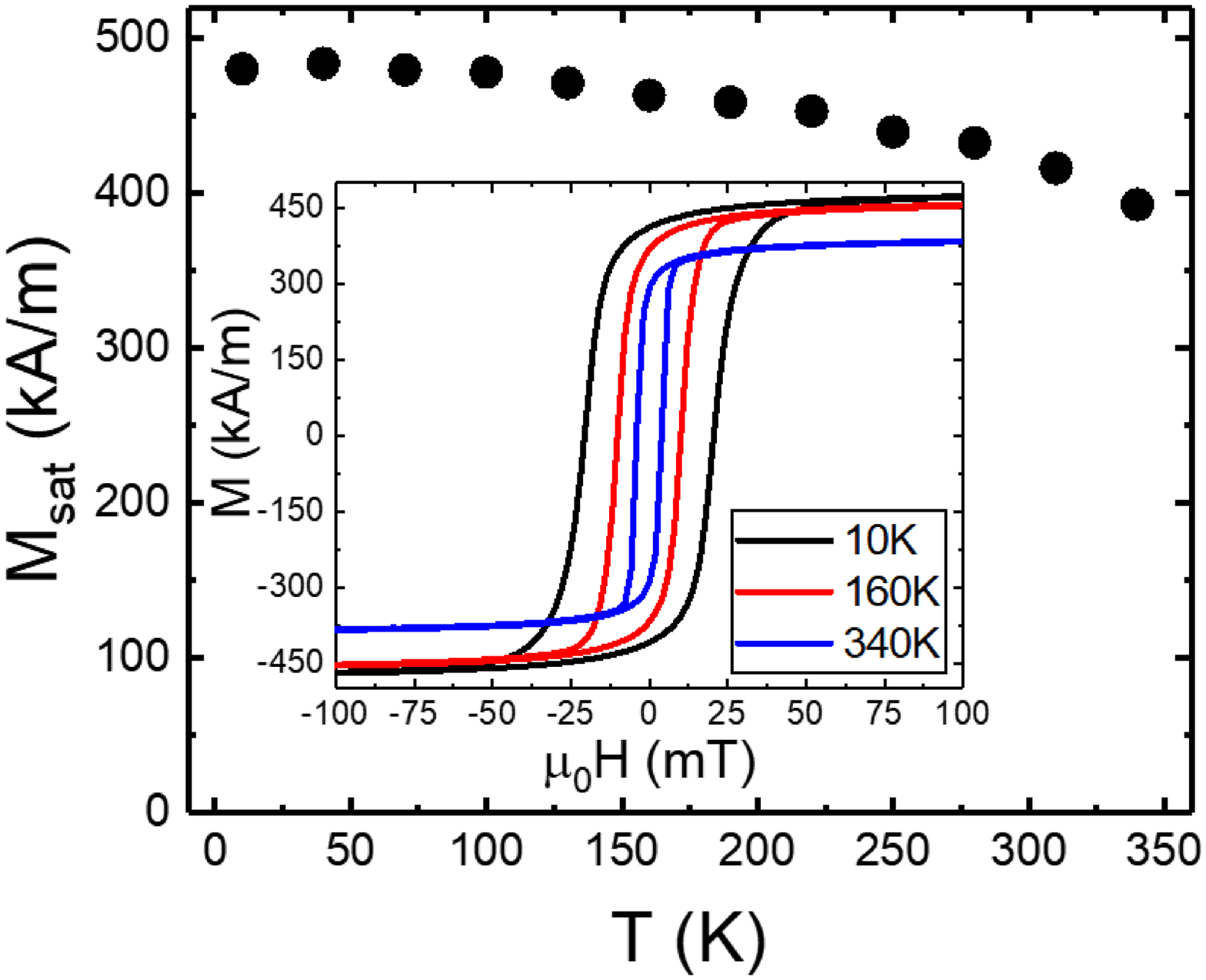}
\caption{Saturation magnetization vs temperature for a Ni(4.0) thin film. \textbf{Inset:} Magnetization field loops at selected temperatures.}
\label{fig:NiTemp}
\centering
\end{figure}

We used our Ni(4.0) sample to compare the temperature dependence of both the hysteresis and of $M_{sat}$ with those of the Ni(2.0)/Ru(2.3)/Ni(2.0) SAF. We measured $M$ vs $H$ loops for this sample from 10 K to 340 K in 30 K steps, as shown in Fig. \ref{fig:NiTemp}. The temperature dependence of $M_{sat}$ is similar for the two samples. In addition, the field offsets between the upsweep and downsweep data (widths of the hysteresis loop) are also similar, confirming that the large openings seen in the Fig. \ref{fig:NiRuTemp} inset are indeed due to the magnetocrystalline and disorder-induced anisotropy of the Ni. The $M_{sat}$ values are higher for Ni than for the SAF, due to the magnetic dead layers at the two Ni/Ru interfaces discussed earlier.

\begin{figure}[!htbp]
\includegraphics[width=\linewidth]{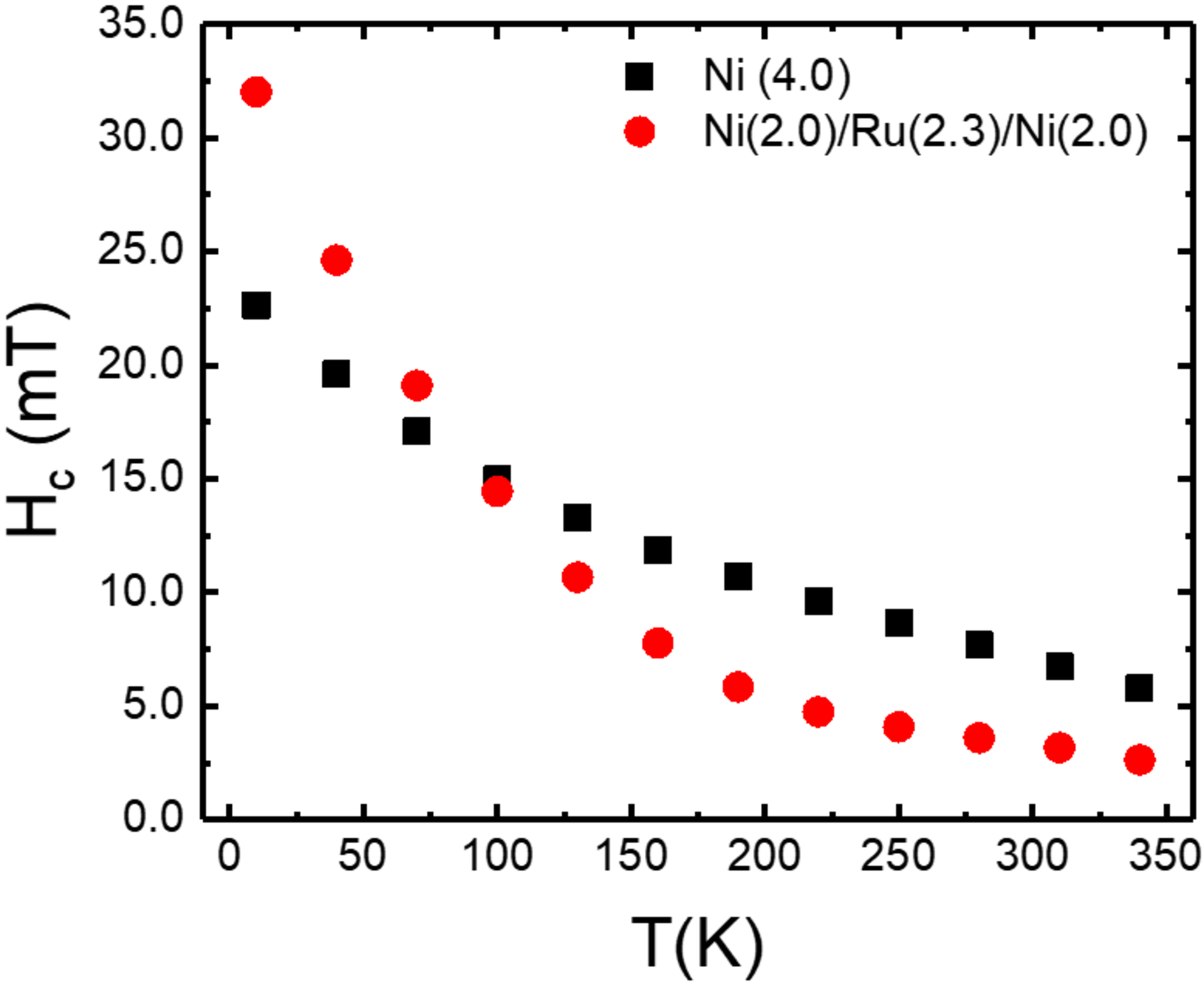}
\caption{Coercivity vs temperature for Ni(4.0) and Ni(2.0)/Ry(2.3)/Ni(2.0) thin films.}
\label{fig:HcCompare}
\centering
\end{figure}

We determined the coercivity of Ni(4.0) by fitting the $M$ vs $H$ loops to the error function. Since the same was not possible for Ni(2.0)/Ru(2.3)/Ni(2.0), we determined the coercivity by taking the average of the coercivity at $M=0$ for upsweep and downsweep of the $M$ vs $H$ curves after fitting a few points around the $M=0$ crossing point to a straight line. Figure \ref{fig:HcCompare} shows the results. At high temperatures, the SAF has lower coercivity than the plain Ni film, as one would expect for a SAF. At lower temperature, however, the opposite is true. That may be a result of the fact that each Ni film in the SAF has a thickness of only 2 nm rather than 4 nm, which makes the comparison somewhat unfair. The coercivity of Ni films increases rapidly as their thickness is reduced, as shown in Fig. \ref{fig:NiSat} of this Supplement.

\begin{figure}[!htbp]
\includegraphics[width=\linewidth]{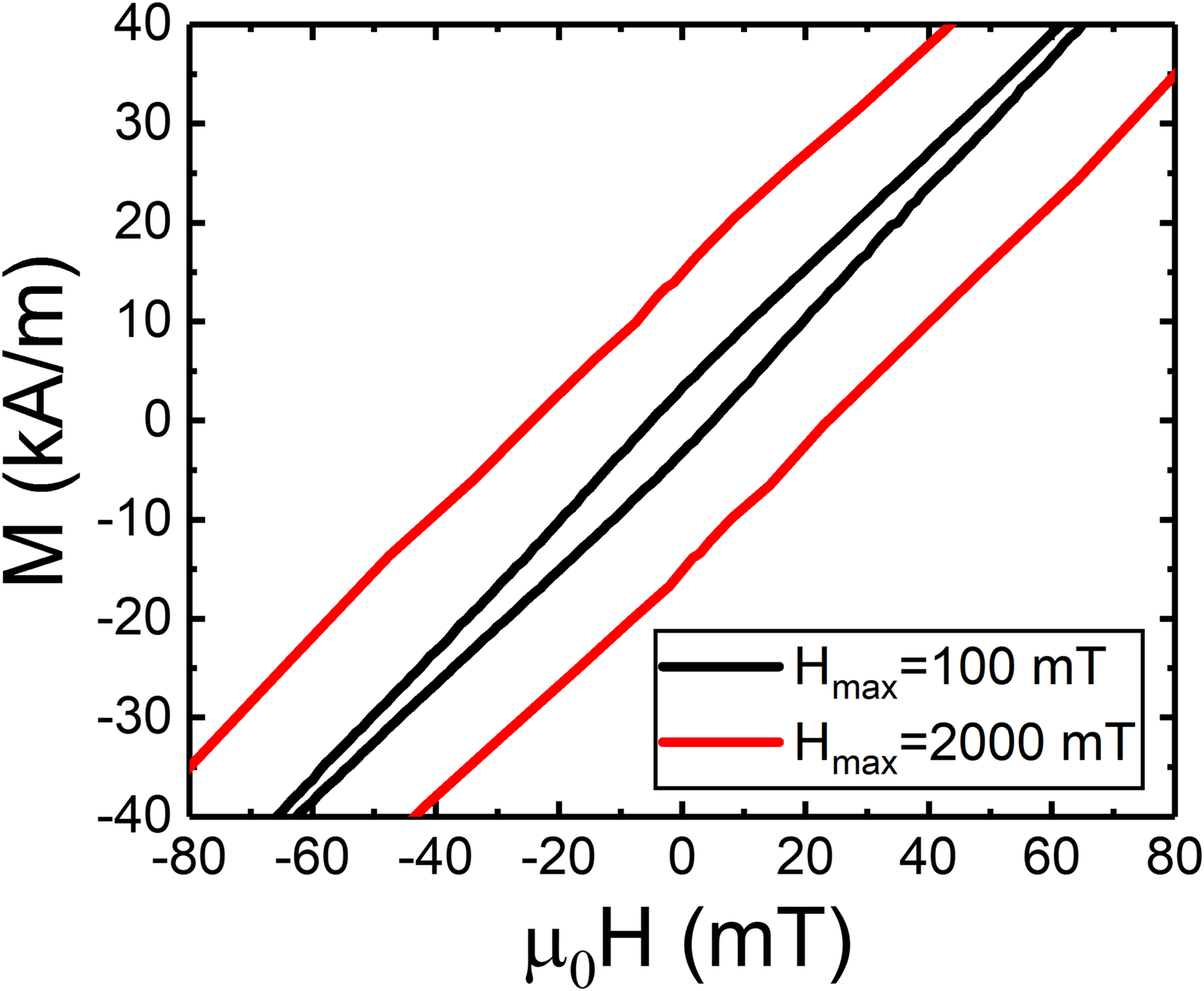}
\caption{Blow-up of magnetization vs field loops Ni(2.0)/Ru(0.9)/Ni(2.0) measured at $T = 10$ K for $H_{max}$ = 100 and 2000 mT.}
\label{fig:Hlow}
\centering
\end{figure}

The inset to Fig. 3 in the main paper shows a Fraunhofer pattern well centered at zero field, indicating that the Ni/Ru/Ni SAF inside the junction has no discernible remanent magnetic moment.  That appears to contradict the remanent magnetization visible in the $M$ vs $H$ curves shown in the inset of Fig. 1.  The difference is due to the different field ranges of the two measurements -- 2 T for the film characterization measurements vs 100 mT for the Josephson junction measurements.  To clarify this issue, we remeasured the $M$ vs $H$ loop for the Ni(2.0)/Ru(0.9)/Ni(2.0) SAF, but this time only up to a field of 100 mT to be consistent with the field range of our $I_c R_N$ measurements. Fig \ref{fig:Hlow} shows a blow-up of both curves. As we can see, the remanent magnetization in the case of low $H_{max}$ is much smaller when compared to that of high $H_{max}$. To estimate the amount of shift this remanent magnetization would produce in our junctions, we modify the expression for the magnetic flux $\Phi$ that appears in Eqn. 2 of the main paper to the following:
\begin{equation}
    \Phi = \mu_0 H (2\lambda_L + d_N + d_F) w + \mu_0 M d_F w
\end{equation}
where M is the net magnetization of the ferromagnetic layers and all other quantities are the same as before. This implies that the magnetization would shift the Fraunhofer curves by a field
\begin{equation}
    H_{shift} = \frac{-M d_F}{(2\lambda_L + d_N + d_F)}
\end{equation}
The $M_{rem}$ in the low $H_{max}$ measurement for Ni(2.0)/Ru(0.9)/Ni(2.0) is estimated to be 3.3 kA/m. Using this value in the above equation, we calculate the shift around zero applied field in the Fraunhofer pattern to be $\mu_0 H_{shift}$ = -0.12 mT. This shift is too small to be visible; hence our Fraunhofers patterns are symmetric around zero field and exhibit no visible hysteresis between upsweep and downsweep.

\bibliography{}